\documentclass[showpacs,preprintnumbers,amsmath,amssymb,10pt]{revtex4}
\usepackage{graphicx,amssymb,amsfonts}
\newcommand{\be}{\begin{equation}}
\newcommand{\ee}{\end{equation}}

\newcommand{\bea}{\begin{eqnarray}}
\newcommand{\eea}{\end{eqnarray}}
\begin{document}
\begin{center}
\vspace{3cm}
\end{center}
\title{\Large \bf Neutral Higgs Sector of the
    MSSM with Explicit $CP$ violation and Non-Holomorphic Soft
    Breaking}
\author{Elif Cincio\~glu$^{a}$, Alper Hayreter$^{b}$, Asl{\i} Sabanc{\i}$^{c}$ and Levent Solmaz$^{a}$}
\affiliation{$^a$ Department of Physics, Bal{\i}kesir University,
Bal{\i}kesir, Turkey, TR10145} \affiliation{$^b$ Department of
Physics, Concordia University, 7141 Sherbrooke West, Montreal,
Quebec, Canada, H4B 1R6} \affiliation{$^c$ Department of Physics and Helsinki Institute of Physics (HIP),
P.O. Box 64, FIN-00014 University of Helsinki, Finland}
\date{\today}
\begin{abstract}
\medskip
Using the effective potential method, we computed one-loop
corrections to the mass matrix of neutral Higgs bosons of the
Non-Holomorphic Supersymmetric Standard Model (NHSSM) with explicit
CP violation, where the radiative corrections due to the quarks and
squarks of the third generation were taken into account.

We observed that the non-holomorphic trilinear couplings  can compete
with the holomorphic ones in CP violating issues for the mass and mixing
of the neutral Higgs bosons.
\end{abstract}
 \maketitle
\section{Introduction}
In the Minimal Supersymmetric Extension of the Standard Model (MSSM)
superpotential and soft breaking terms are generally considered as
holomorphic functions. While the holomorphicity of the
superpotential is obligatory for the MSSM, more generalized versions
of the  model may also include R-parity violating terms and/or
non-holomorphic structures in the soft breaking sector of the
theory \cite{jack1,jack2}.

Contrary to the SM where a unique  Higgs boson resides,
supersymmetric models predict extra Higgs bosons via introducing
different Higgs doublets. With a number of supersymmetric models and
various extensions, experimental verification of Higgs bosons became
one of the main objectives of the current colliders such as Tevatron
\cite{tevatron} and LHC \cite{LHC}. Compared with the SM, the
allowed mass range of the lightest Higgs in the MSSM is somewhat
more constrained. Indeed, the predictions related to the mass of the
lightest Higgs in the MSSM give an upper bound $m_h\sim$ 130 GeV
\cite{Higgs bound} which can be fairly relaxed by certain extensions
of the theory. In this sense, the Non-Holomorphic Supersymmetric
Standard Model (NHSSM) requires the presence of the additional soft
breaking parameters that can shift the upper bound to a certain
extend \cite{bizim} which may be required in further Higgs searches.

In addition to give a relaxation to the upper bound of lightest
Higgs mass, these additional non-holomorphic soft breaking terms
ensure extra degrees of freedom where, for instance, CP violating
terms of the MSSM may get into trouble. It is explicitly shown in
\cite{Gavela:1994dt} that the amount of CP violation present in the
SM is not adequate to explain the observed baryon asymmetry in the
universe, whereas, supersymmetric models offer novel sources of CP
violating terms  and especially  Higgs interactions can play a key
role in mediating CP violation. This issue is deeply scanned in the
MSSM \cite{wagner}. But, this should also be probed for the
extensions of the minimal model where additional sources of CP
violating terms exist. In this respect, the NHSSM is an interesting
model with extra sources of CP violating terms in the soft breaking
part of its Lagrangian. However before doing this, precise
predictions are required for the Higgs sector of the model which
does not exist in the literature.

Hence, in this work, our interest focused on the neutral Higgs
sector of the Non-Holomorphic Supersymmetric Standard Model (NHSSM)
with R-parity conservation and explicit CP violation. We assumed
that CP is explicitly violated in the Higgs sector of the NHSSM and
looked for its impact on the mass and mixing of the neutral Higgs
bosons, which may be important for the near future. The possibility
of non-holomorphic structures in the soft breaking is realized in
the literature. For a detailed list of issues ranging from
$b\rightarrow\,s\gamma$ decay to the Renormalization Group Equations
(RGEs) of the non-holomorphic supersymmetric model, we refer to
\cite{Hetherington:2001bk}, a possible explanation for the source of
the NH structures can be found in  \cite{Haber:2007dj}.

The rest of the paper is as follows: In the following section we
first described the basic low energy structure of the NHSSM.
Analytical results for the mass matrix of the neutral Higgs bosons
of the NHSSM are derived in the same section where the one-loop CP
violating effective potential is calculated by considering only top
and bottom sectors. Section III is devoted to numerical analysis
where the impact of non-holomorphic trilinear terms are
investigated. And we concluded in section IV.
\section{NHSSM}
In general the supersymmetry (SUSY) breaking sector is parameterized
via holomorphic operators which must be soft  $i.e.$ the quadratic
divergences must not be regenerated \cite{Chung:2003fi}. However,
the MSSM can be extended via introducing new soft operators
including R-violating and/or non-holomorphic terms in the soft
breaking sector of the theory. In this sense there are different
non-holomorphic models based on different approaches  (see
\cite{jack1} and \cite{Hetherington:2001bk}). Here we follow the
easiest path in which R-parity violating terms are ignored and the
problematic Higgsino mass term ($\mu$) is absent in the
superpotential. Under these assumptions, the NH version of the
minimal supersymmetric model can be described by the superpotential
\begin{eqnarray}
\label{sup} \widehat{W} =  \widehat{Q} \cdot \widehat{H}_u {\bf Y_u}
\widehat{U} -\widehat{Q} \cdot \widehat{H}_d  {\bf Y_d} \widehat{D}
-\widehat{L} \cdot \widehat{H}_d {\bf Y_e} \widehat{E}
\end{eqnarray}
where  our conventions are such that, for instance, $\widehat{Q}
\cdot \widehat{H}_u \equiv \widehat{Q}^{T} (i \sigma_2)
\widehat{H}_u = \epsilon_{i j} \widehat{Q}^i \widehat{H}^j_u$ with
$\epsilon_{12}=-\epsilon_{21}=1$.
In the MSSM the breakdown of
supersymmetry is parameterized by a number of holomorphic soft
operators \cite{Chung:2003fi}
\begin{eqnarray}
\label{soft} -{\cal{L}}_{soft}&=& \tilde{Q}^{\dagger} {\bf
m}_{\tilde{Q}}^2 \tilde{Q} + \tilde{U}^{\dagger} {\bf
m}_{\tilde{U}}^2 \tilde{U} + \tilde{D}^{\dagger} {\bf
m}_{\tilde{D}}^2 \tilde{D} + \tilde{L}^{\dagger} {\bf
m}_{\tilde{L}}^2 \tilde{L} + \tilde{E}^{\dagger} {\bf
m}_{\tilde{E}}^2 \tilde{E} +\frac{1}{2} \left(M_3
\lambda_{\tilde{g}}^{a} \lambda_{\tilde{g}}^{a} +
 M_2 \lambda_{\tilde{W}}^{i} \lambda_{\tilde{W}}^{i} +
M_1 \lambda_{\tilde{B}} \lambda_{\tilde{B}} + \mbox{h.c.} \right)\nonumber\\
&+& m_{H_u}^2 H_{u}^{\dagger} H_u + m_{H_d}^2 H_{d}^{\dagger} H_d +
\left( m^2_3 H_{u}\cdot H_d + \mbox{h.c.}\right)+\left(
\tilde{Q}\cdot {H}_u {\bf Y_u^A} \tilde{U} -\tilde{Q}\cdot {H}_d
{\bf Y_d^A} \tilde{D}
 - \tilde{L}\cdot {H}_d {\bf Y_e^A} \tilde{E} +
 \mbox{h.c.}\right).\,\,\,\,
\end{eqnarray}
Here ${\bf m}_{\tilde{Q},\cdots,\tilde{E}}^2$ are the soft
mass-squareds of the scalar fermions, ${\bf Y_{u,d,e}^A}$ are their
associated holomorphic trilinear couplings, and finally, $M_1, M_2,
M_3$ are, respectively, the masses of hypercharge, isospin and color
gauginos. For the description of the Higgs sector  soft masses
$m_{H_u}^2$, $m_{H_d}^2$ and $m^2_3$ are used.

In the MSSM one can introduce the CP violation  through the Higgs
superpotential and the soft supersymmetry breaking terms, however,
as has been shown explicitly in \cite{Girardello:1981wz,poppitz}, in
supersymmetric theories which do not have pure gauge singlets in
their particle spectrum, the holomorphic supersymmetry breaking
terms do not necessarily represent the most general set of
soft-breaking operators. Indeed, for instance, the MSSM spectrum
does not consist of any gauge singlet superfield, and thus, its soft
breaking sector must necessarily include the following soft breaking
terms
\begin{eqnarray}
{\cal{L}}_{soft}^{\prime} =\mu^\prime {\tilde{H}_u}\cdot
{\tilde{H}_d} +\tilde{Q}~{H}_d^{C} {\bf {Y^\prime}_u^A} \tilde{U}+
\tilde{Q}~ {H}_u^{C} {\bf {Y^\prime}_d^A} \tilde{D}
 + \tilde{L}~{H}_u^{C} {\bf {Y^\prime}_e^A} \tilde{E} + \mbox{h.c.}
\end{eqnarray}
in addition to those in (\ref{soft}). Here ${\bf
{Y^\prime}_{u,d,e}^A}$ are non-holomorphic trilinear couplings which
do not need to bear any relationship to the holomorphic ones ${\bf {
Y}_{u,d,e}^A}$ in (\ref{soft}). The only exception to this  can be
imagined as a unique common term at very high scales but even in
that case, due to renormalization group running effects it is good
to assume those new trilinear couplings to be completely different
from the ordinary ones. Since these non-holomorphic couplings are
perfectly soft they must be taken into account when confronting the
MSSM predictions with experimental data.

For possible variants of the non-holomorphic model notice that the
original $\mu$ term can be protected in the superpotential, in this
case, the soft breaking $\mu^\prime$ can stand alone or replaced
with $\mu^\prime-\mu$, then the $m^2_3$ term of the soft sector can
be written as $m^2_3=B(\mu-\mu^\prime)$. But we have chosen  to deal
with only one $\mu$ parameter for which the prime symbol will be
dropped from now on.
\subsection{Analytical Results for the Neutral Higgs Bosons}
As in the MSSM, the CP violation mixes  neutral Higgs bosons and
hence they will be depicted as physical mass eigenstates $h_1, h_2$
and $h_3$. The classical potential for the neutral Higgs fields can
be written for the NHSSM like in the MSSM \cite{Martin:1997ns} as
follows
\begin{eqnarray}
\label{higgstree} V&=&m^2_{H_u} |H^0_u|^2+m^2_{H_d} |H^0_d|^2-(m^2_3
H^0_u H^0_d + c.c.)+\frac{{\hat{g}}^{2}}{8}(|H^0_u|^2-|H^0_d|^2)^2+
\Delta V
\end{eqnarray}
Here ${\hat{g}}^{2}=g_2^2 + g_Y^2$ and $\Delta V$ refers to loop
corrections which will be computed using the effective potential
formalism and  the symbols $g_2$ and $g_Y$ stand for the SU$(2)$ and
U$(1)$ gauge couplings respectively. Contrary to the MSSM, in
(\ref{higgstree}) $|\mu|^2$ contributions coming from $F-$terms are
absent which stem from the superpotential (\ref{sup}). In
(\ref{higgstree}) we allowed $m^2_3$ to be complex by assuming:
\begin{eqnarray}
m^2_3 =|m^2_3|  e^{i \Phi}\,.
\end{eqnarray}
This phase $\Phi$ can be set to zero at the tree level, but due to
loop corrections it should be protected. Our results for Higgs
masses will depend on this and another phase coming from one
of the Higgs doublets. Now, the neutral components of the Higgs
doublets can be expanded around their vacuum expectation values
(vevs) as
\begin{eqnarray}
\label{higgsexp}  H^0_d &=& \frac{1}{\sqrt{2}}\left( v_d + \phi_d +
i a_1\right),\:\:\:\:  H^0_u  = \frac{e^{i \theta}}{\sqrt{2}}\left(
v_u + \phi_u + i a_2\right)
\end{eqnarray}
in which $v^2\equiv v_u^2+v_d^2=(246\,{\rm GeV})^2$. Using the ratio
of vevs we define $\tan\beta=v_u/v_d$. In the above expression, a
phase shift $e^{i\theta}$ is attached to neutral part of the up
Higgs doublet $H^0_u$ and this phase should be fixed by true vacuum
conditions considering loop effects (see \cite{higgs1,higgs2,higgs3,Demir}
for details).

It is  important to emphasize that without loop corrections the tree
level Higgs potential (\ref{higgstree}) predicts the lightest Higgs
boson to be lighter than the $Z$ boson. Hence, as in the MSSM,
sizeable radiative corrections are also needed in the NHSSM to
satisfy the LEP bound of $m_h\sim 114\, {\rm GeV}$. In this work we
did not consider LEP excess events which indicate on a possibility
of even a lighter Higgs \cite{Barate:2003sz}, however, the excess
events may still remain as another issue to be considered within the
context of the NHSSM.

In the MSSM, the radiative corrections \cite{higgs1,higgs2} are
dominated by loops of the top (s)quark, and to a lesser extent, by
those of the bottom (s)quark, tau (s)lepton, charginos and
neutralinos \cite{higgs3}. A particularly useful framework for
computing the radiative corrections in the Higgs sector is effective
potential approach \cite{Coleman:1973jx}. At the one-loop level we
can write the contributions of all the relevant particles (coupled
to Higgs bosons) as
\begin{eqnarray}
\Delta V=\frac{1}{64\,\pi^2} Str\left[\mathcal{M}^4\left(\ln
\frac{\mathcal{M}^2}{\Lambda^2}-\frac{3}{2}\right)\right]
\end{eqnarray}
Here, $\Lambda$ is the renormalization scale and $\mathcal{M}$ is
the field-dependent mass matrix of quarks and squarks with  overall
factors -12 and 6, respectively. The additional contributions coming
from charginos, neutralinos, etc.  are ignored in this work.

To proceed, the relevant fermion masses should be stated in
a field dependent manner as in the background formed by the neutral
components of the Higgs fields, for instance, the squared-mass of
bottom and top quarks are given by
\begin{eqnarray}
\label{topmass} m^2_b =|h_b|^2 \mid H^0_b\mid^2,\,\,\,\,\,\,\,\,
m^2_t &=& |h_t|^2 \mid H^0_u\mid^2
\end{eqnarray}
and those of the scalar quarks are
\begin{eqnarray}
\label{sbotmass} M^2_{\tilde{b}} = \left( \begin{array}{c c}
m^2_{\tilde{t}_L} + m^2_b +\frac{1}{12} (3g_2^2+g_Y^2)(|H_u^0|^2 - |
H_d^0|^2) &
h_b\left(A_b {H^0_d}-{A^\prime_b}^* {H^0_u}^*\right)\\
h^*_b\left(A^*_b {H^0_d}^*-{A^\prime_b} {H^0_u}\right) &
m^2_{\tilde{b}_R}+m^2_b+\frac{1}{6} g_Y^2(|H_u^0|^2
-|H_d^0|^2)\end{array}\right),
\end{eqnarray}
\begin{eqnarray}
\label{stopmass} M^2_{\tilde{t}} = \left(\begin{array}{c c}
m^2_{\tilde{t}_L} + m^2_t- \frac{1}{12} (3g_2^2-g_Y^2) (|H_u^0|^2 -
|H_d^0|^2) &
h_t\left( A_t {H^0_u}-{A^\prime_t}^* {H^0_d}^*\right)\\
h^*_t\left(A^*_t {H^0_u}^*-{A^\prime_t} {H^0_d}\right) &
m^2_{\tilde{t}_R} + m^2_t-\frac{1}{3}g_Y^2 (|H_u^0|^2
-|H_d^0|^2)\end{array}\right).
\end{eqnarray}
In writing the squark mass-squared matrices we have introduced some
notationally simplifying definitions such as $({\bf
m}_{\tilde{Q}}^2)_{33} \equiv m_{\tilde{t}_L}^2$ and $({\bf
{Y^\prime}_{d}^A})_{33} \equiv h_b { A^\prime}_b$. In
(\ref{sbotmass}) and (\ref{stopmass}), the main effect of
non-holomorphic trilinear couplings  is to replace the $\mu$
parameter in the holomorphic MSSM in a flavor-dependent way and this
shift alone tells us that the $\mu$ parameter seen by Higgsinos is
completely different than what is felt by the scalar fermions. It is
possible to back-transform $A^\prime_t,A^\prime_b\rightarrow\mu$ to
obtain the MSSM results, but the reverse is not true. In other
words, the indirect relation between scalar fermions and charginos
or neutralinos over the $\mu$ parameter is completely vanished in
this version of the NHSSM. From now on, bounds on the $\mu$
parameter ($i.e.$ obtained from charginos) have no restriction on
scalar fermions anymore. Besides this the mass of the Higgsinos is
the same with the MSSM if we assume $\mu$ as an input parameter.

In this work, rather than providing a general analysis of ${\bf
{Y^\prime}_{u,d,e}^A}$ in regard to MSSM phenomenology (see
\cite{Cakir:2006ut} attempts in this direction), we will focus
mainly on their influence on Higgs-fermion-fermion couplings
(especially for $h_i\rightarrow\bar{b}b$ decay) in order to
determine their distinctive features and observability in collider
experiments. Concerning this class of observables, the primary
objective would be to determine sensitivities of Higgs boson masses
and mixings to the non-holomorphic couplings ${\bf
{Y^\prime}_{u,d}^A}$. For this purpose, to leading order, it
suffices to consider only the top and bottom quark sector.

Now, for later convenience, we introduce $\Sigma$ and $\Delta$
symbols such that (s)top and (s)bottom  mass eigenvalues can be
written simply as \be \label{stop-mass}
m_{\tilde{t}_{1,2}}^2=\left(\Sigma_T\mp\sqrt{\Delta_T}\right)/4~,\,\,\,\,\,\,m_{\tilde{b}_{1,2}}^2=\left(\Sigma_B\mp\sqrt{\Delta_B}\right)/4
\ee satisfying hierarchical order $m_{\tilde{f}_{1}}^2 <
m_{\tilde{f}_{2}}^2$ with $f=b,t$. Of course, the sfermion masses
appearing in (\ref{stop-mass}) correspond to  field dependent
$m_{\tilde{f}_{1,2}}^2$  evaluated in the electroweak vacuum. The
explicit form of  our definitions can be found in the Appendix.

The calculation proceeds by plugging the field-dependent eigenvalues
into the potential. The mass matrix of the Higgs bosons is given by
the second derivatives of the potential (at vanishing external
momentum). For this aim, the minimum of the potential should be
obtained, which can be extracted from the first derivatives of the
potential V (\ref{higgstree}). In turn, $m^2_{H_u}, m^2_{H_d}$ and
$m^2_3$ can be expressed in terms of functions of the parameters
appearing in the loop-corrected Higgs potential. We define \be
\mathcal{T}_i=\partial V/\partial \Psi_i ~,~~~~~~
\mathcal{M}^2_{ij}=\partial^2 V/\partial\Psi_i
\partial\Psi_j\ee for the first and second derivatives
respectively with $\Psi_i,\Psi_j=\phi_u,\phi_d, a_1, a_2$ and both
$\mathcal{T}$ and $\mathcal{M}^2$ should be evaluated at vacuum
conditions $\Psi_i,\Psi_j=0$. Among the stationary relations
$\mathcal{T}_{\phi_u}$ and $\mathcal{T}_{\phi_d}$ are linearly
independent. But $\mathcal{T}_{a_1}$ and $\mathcal{T}_{a_2}$ can be
expressed in terms of each other. Hence we can express $m^2_3$ as
follows \be \label{m32}
\frac{3}{16\pi^2}\csc[\theta+\Phi]\bigg\{\frac{\mathcal{I}_b\,|h_b|^2}{\sqrt{\Delta_B}}
\bigg[\sqrt{\Delta_B}+2m^2_{\tilde{b}_1}\ln\frac{m^2_{\tilde{b}_1}}{\Lambda^2}-2m^2_{\tilde{b}_2}\ln\frac{m^2_{\tilde{b}_2}}{\Lambda^2}\bigg]
+\frac{\mathcal{I}_t\,|h_t|^2}{\sqrt{\Delta_T}}\bigg[\sqrt{\Delta_T}+2m^2_{\tilde{t}_1}\ln\frac{m^2_{\tilde{t}_1}}{\Lambda^2}-2m^2_{\tilde{t}_2}\ln\frac{m^2_{\tilde{t}_2}}{\Lambda^2}\bigg]\bigg\}
\ee
 In this equation $\mathcal{I}_{f}=(A_f A^\prime_f e^{i\theta})$
describes the amount of CP violation in the sfermion mass matrices.
Notice that the combination of phases $\theta+\Phi$ is re-phasing
invariant and validity of (\ref{m32}) can be checked from
\cite{Choi:2000wz} in the
$(A^\prime_{t},A^\prime_{b})\rightarrow\mu$ limit. During the
numerical analysis we fixed $\theta=-\pi/2$ and determined $\Phi$ in
accordance with the input parameters.

After obtaining true tadpoles correctly, Higgs boson mass-squared
matrix (${M}^2_{ij}$) is acquired in the base of $\lbrace
\phi_u,\phi_d, a_1, a_2 \rbrace$ in the form of a symmetrical
$4\times4$ matrix. The eigenvalues of this symmetric mass-squared
matrix correspond to a massless Goldstone boson and three physical
neutral Higgses ($m^2_{h_1},m^2_{h_2}$ and $m^2_{h_3}$). They can be
used for numerical purposes, but for analytical purposes it is
useful to perform the following unitary transformation;
\begin{eqnarray}
\label{ma-sq} {\mathcal{M}}^{2} =\mathcal{S}^T {M}^{2}
\mathcal{S}\rm{~~~~where~~}{\mathcal{S}}= \left(\begin{array}{cc}
1& 0 \\
 0& \eta \\
\end{array}\right)\rm{~~~~and~~}
\eta=
 \left(\begin{array}{cc}
s_\beta&c_\beta \\
c_\beta&-s_\beta \\
\end{array}\right).
\end{eqnarray}
This transformation  allows us to redefine $\mathcal{M}^2$ as a
symmetric $3\times3$ matrix in the basis $\lbrace \phi_u, \phi_d, a
\rbrace$ where $a$ is defined as a linear combination of $a_1$ and
$a_2$ ($a=\sin\beta\,a_1+\cos\beta\,a_2$). For instance
$\mathcal{M}^2_{33}$ component of the redefined $\mathcal{M}^2$
matrix becomes,
 \bea
 \label{m33}
&&\mathcal{M}^2_{33}={m^2_3}\frac{v^2\cos(\theta+\Phi)}{{v_d}{v_u}}+\frac{3|h_t|^2v^2\mathcal{R}_t}{32\pi^2{v_d}{v_u}}\ln\frac{m^2_{\tilde{t}_1}m^2_{\tilde{t}_2}}{\Lambda^4}+\frac{3|h_b|^2v^2\mathcal{R}_b}{32\pi^2{v_d}{v_u}}\ln\frac{m^2_{\tilde{b}_1}m^2_{\tilde{b}_2}}{\Lambda^4}\nonumber\\
&&+\frac{3|h_t|^2v^2\Sigma_T\left(8|h_t|^2{v_d}{v_u}\mathcal{I}_t^2-\Delta_T\mathcal{R}_t\right)}{32\pi^2{v_d}{v_u}\Delta_T^{3/2}}\ln\frac{m^2_{\tilde{t}_1}}{m^2_{\tilde{t}_2}}+\frac{3|h_b|^2v^2\Sigma_B\left(8|h_b|^2{v_d}{v_u}\mathcal{I}_b^2-\Delta_B\mathcal{R}_b\right)}{32\pi^2{v_d}{v_u}\Delta_B^{3/2}}\ln\frac{m^2_{\tilde{b}_1}}{m^2_{\tilde{b}_2}}\nonumber\\
&&+\frac{3}{32\pi^2v^2{v_d}{v_u}\Delta_B\Delta_T}\left\{16|h_b|^4v^4{v_d}{v_u}\Delta_T \mathcal{I}_b^2-\Delta_B \left(({v^4_d}+{v^4_u}) \Delta_T (2|h_b|^2\mathcal{R}_b+2|h_t|^2\mathcal{R}_t) \right.\right.\nonumber\\
&&\,\,\,\,\,\,\,\,\,\,\,\,\,\,\,\,\,\,\,\,\,\,\,\,\,\,\,\,\,\,\,\,\,\,\,\,\,\,\,\,\,\,\,\,\,\,\,\,\,\,\,\,\,\,\,\,\left.\left.+4{v_d}{v_u}\left(-4|h_t|^4v^4\mathcal{I}_t^2+{v_d}{v_u}\Delta_T\left(|h_b|^2\mathcal{R}_b+|h_t|^2\mathcal{R}_t\right)\right)\right)\right\}
\eea The explicit form of the symbols given here can be read from
the Appendix. We refer to the same place for the remaining five
entries of the symmetric $\mathcal{M}$ matrix.
\section{Numerical Analysis}
In this part, based on our analytical results, our aim is to show
how the interplay of the non-holomorphic couplings with holomorphic
ones can change the mass and the mixing of the neutral Higgs bosons.

During the  analysis, to respect the collider bounds, we require the
scalar fermion masses satisfying $m_{\tilde{f}}>100 {\rm{~GeV}}$ and
generally our results cover the LEP bound $m_{h_1}\sim\,114$ GeV.
 Our basic input parameters are $M_A,m_{\tilde Q},m_{\tilde
U},m_{\tilde D},A_t,A_b,A^\prime_t,A^\prime_b$ and $\tan\beta$.
During the analysis we fixed $\theta=-\pi/2$, $\Lambda=0.5$ TeV and
determined  $\Phi$ in accordance with the input parameters. The true
phase of CP violation can be defined as $\theta_{eff}=arg(A_f
A^\prime_f e^{i\theta})$ for top and bottom sectors and hence
variation of  $\Phi$ is never presented. Instead we concentrated on
the trilinear couplings and looked mainly for the mass and the
mixings of the Higgs bosons under CP violating non-holomorphic
trilinear couplings.

 In the numerical analysis we have
taken $M_A$ as one of the input parameters and forced
$M_A={\mathcal{M}}_{33}$, which is slightly different from  the
selection of ref. \cite{Choi:2000wz}. Alternatively, mass of the
charged Higgs boson can be used as an input parameter as it is
usually done in the MSSM literature ($i.e.$ see \cite{wagner}). To
make the neutral Higgs mass matrix diagonal, we also defined an
orthogonal matrix O such that \be
{\rm{diag}}(m^2_{h_1},m^2_{h_2},m^2_{h_3})={O}^T {\mathcal{M}}^{2}
{O}, \ee
then, an additional  parameter can be defined to describe
the CP composition of the neutral Higgs bosons as \cite{Choi:2000wz}
 \be
\alpha_i={\rm\,min}\bigg(\frac{|O_{i3}|}{\sqrt{|O_{i1}|^2+|O_{i2}|^2}},\frac{\sqrt{|O_{i1}|^2+|O_{i2}|^2}}{|O_{i3}|}\bigg)
\ee With these definitions CP composition of neutral Higgses
under the influence of non-holomorphic trilinear couplings can be
presented. Additionally, we have selected the decay width of
neutral Higgs bosons into $\bar{b}\,b$ as a testing ground. Of
course ($h_i\rightarrow\bar{b}b$) has massive background, but
according to the SM up to $M_H\sim130$ GeV this channel is the
dominant one. So we have chosen this channel for its simplicity to
show the impact of the phases. Notice in the SM that the width of
this decay is 0.0035 to 0.086 GeV for $m_h$ between 120 - 160 GeV
\cite{Ham:2007mt}. But the NHSSM has potential to change it sizably,
as we will see. The partial decay width of a neutral Higgs boson
$h_i$ into a pair of bottom and anti-bottom quarks is given as
\cite{Ham:2007mt} \be \Gamma(h_i\rightarrow\bar{b}\,b)= \frac{3
g^2_2 m^2_b m_{h_i}} {32 \pi m^2_W} \sqrt{1
-\frac{m^2_b}{m^2_{h_i}}}\bigg[\frac{O^2_{i1}}{\cos^2 \beta} (1
-\frac{m^2_b}{m^2_{h_i}}) + \tan^2\beta\,O^2_{i3}\bigg].\ee
\begin{figure}[h!]
\begin{center}
\includegraphics[scale=1,height=13cm,angle=90]{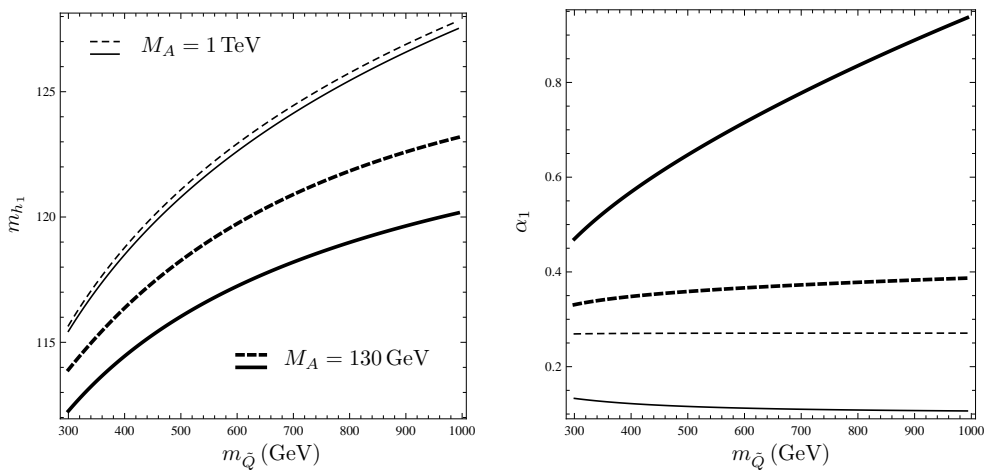}
\end{center}
\vspace{-0.4cm} \caption{The lightest neutral Higgs mass $m_{h_1}$
(left panel) and its CP violating parameter $\alpha_1$ (right panel)
of the MSSM (solid lines) and of the NHSSM (dashed lines), against
$m_{\tilde Q}$ for $M_A$=1 TeV (thick lines) and $M_A$=130 GeV (thin
lines). Inputs:  $\mu=500$ GeV,  $A^\prime_t=5 m_{\tilde Q}$,
$A^\prime_b$ is fixed at 0, $m_{\tilde Q}=m_{\tilde U}=m_{\tilde
D}$, $A_t=A_b=2 m_{\tilde Q}$ and $m_{\tilde Q}$ scans from 0.3 TeV
to 1 TeV, $\tan\beta$ is fixed at 10. } \label{fig3}
\end{figure}

Now let us present our numerical results. We start with the
difference of $m_{h_1}$ in the MSSM and in the NHSSM for CP
violating case (see \cite{bizim} for CP conserving case).  For the
model under concern, the $\mu$ parameter of the NHSSM is responsible
for higgsino masses as given in the soft breaking sector
(\ref{soft}), and can be bounded from chargino and neutralino masses
as in the MSSM. However in the NHSSM, $\mu$ does not exist in the
sfermions, so whenever we assume $A^\prime_f\neq\mu$ we are
considering the NHSSM. Thus for the differently selected $\mu$ and
$A_t^{\prime}$ values model dependent effects can be observed on the
mass and the mixing of the Higgs bosons.

 In this sense, Fig. \ref{fig3} depicts
the mentioned  differences  of the MSSM with fixed $\mu$ and of the
NHSSM with different $A^{\prime}_t$ and $A^{\prime}_b$ values. The
mass difference, as can be seen from the left panel of the figure,
can be around few MeV  or a few GeV  and increases as the squark
mass increases. This mass difference is determined by the magnitude
of $M_A$ and is  more visible when $M_A$ is close to $m_{h_1}$. A
similar observation can be extracted from the $CP$ violating
parameter of the lightest Higgs boson (right panel). Again when
$M_A\sim\,m_{h_1}$, this parameter shows that the $CP$ composition
of the lightest Higgs can be enhanced as should be expected in the
MSSM \cite{wagner}. In comparison to the MSSM, the CP violating
parameter of the NHSSM can be smaller or larger than that of the
MSSM predictions, which is again determined by $M_A$. For instance,
for $M_A=130$ GeV the $CP$  violating parameter takes
$\alpha^{\rm{NH}}_1\sim\,0.4$ which is approximately half of the
MSSM's prediction but for $M_A=1$ TeV we observe
$\alpha^{\rm{NH}}_1\sim\,0.25$  which is approximately two times
larger than the  prediction of the MSSM. Notice that using this
parameter one can determine whether $h_1$  will behave similar to
the SM's Higgs boson or not. In sum, the effects of the extra
parameter of the  NHSSM ($A^\prime_t$) is more visible when $M_A$ is
close to $m_h$. This is something predictable because in this case
large mixing occurs between the would-be CP-odd and CP-even Higgs
bosons. On the other hand, when  $M_A$ is large it may be hard to
effect the mass of the lightest Higgs boson with the NH terms. But
even in this case its coupling could be very different from the
prediction of the MSSM, which can be read from the right panel of
the figure.

In the same figure, the relaxation on the bound of sfermion masses
can also be deduced for the NHSSM, $i.e.$ as can be seen from the
left panel of Fig. \ref{fig3}, for $m_{h_1}>115$ GeV and $M_A=130$
GeV the MSSM demands $m_{\rm squarks}>450$ GeV, but in the NHSSM
this bound relaxes to $m_{\rm squarks}>350$ GeV, for the selected
range of parameters.

\begin{figure}[h!]
\begin{center}
\includegraphics[scale=1,height=17cm,angle=90]{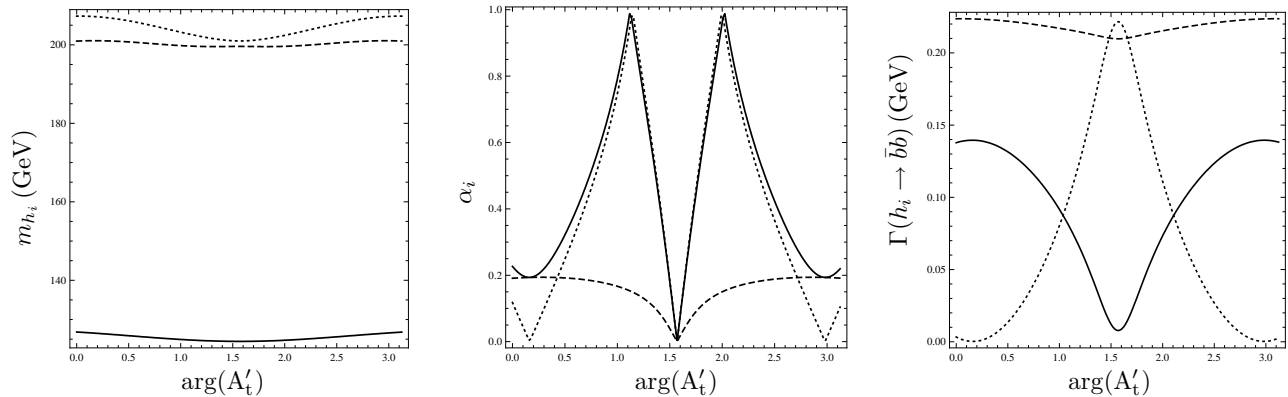}
\end{center}
\vspace{-0.4cm} \caption{Masses of all  neutral Higgs bosons
$m_{h_{1,2,3}}$ (left), their CP-violating mixing angles
$\alpha_{1,2,3}$ (center) and their decay widths into $\bar{b}b$
pair (right) versus the argument of the top trilinear coupling
$\rm{arg}(A^\prime_t)$. The dimensionful terms are given in GeV.
Inputs: $\tan\beta=10$, $m_{\tilde Q}=m_{\tilde U}=m_{\tilde D}=1$
TeV and $A_t=A_b=|A^{\prime}_t|=A^{\prime}_b =2\, m_{\tilde Q}$,
$M_A=200$ GeV. Note that for this and latter figures line formats
are as follows, solid line ($h_1$), dotted line ($h_2$) and dashed
line ($h_3$).} \label{fig5}
\end{figure}

In Fig. \ref{fig5} we present the phase dependencies of the masses
of neutral Higgs bosons $h_1,h_2$ and $h_3$ in left panel, CP
violating parameters ($\alpha_i$) for each of the mentioned bosons
in middle panel and the corresponding decay widths
$\Gamma(h_i\rightarrow\bar{b}b)$ in right panel, against varying
phase of the non-holomorphic trilinear coupling $A^\prime_t$. As can
be seen from the first panel of  Fig. \ref{fig5} all of neutral
Higgs bosons  are sensitive to the phase of the non-holomorphic
trilinear coupling $A^\prime_t$, with varying order. As a result of
this phase the lightest Higgs boson can be made completely CP-odd or
CP-even. One can easily recognize from the middle panel of Fig.
\ref{fig5} that CP violating mixing angles ($\alpha_i$)  are
suppressed for $\rm{arg}(A^\prime_t)\sim \pi/2$ and have a sharp
maximum value for $\rm{arg}(A^\prime_t)\sim \pi/3$ and $\sim
2\pi/3$. Additionally the partial decay widths of the neutral Higgs
bosons are very sensitive to the phase of $A^\prime_t$, can exceed
the prediction of the SM for each of the bosons, $i.e.$
$\Gamma(h_1\rightarrow\bar{b}\,b)\leq0.14$ GeV is possible in the
NHSSM as can be seen from the right panel of Fig. \ref{fig5}.
\begin{figure}[h!]
\begin{center}
\includegraphics[scale=1,height=17cm,angle=90]{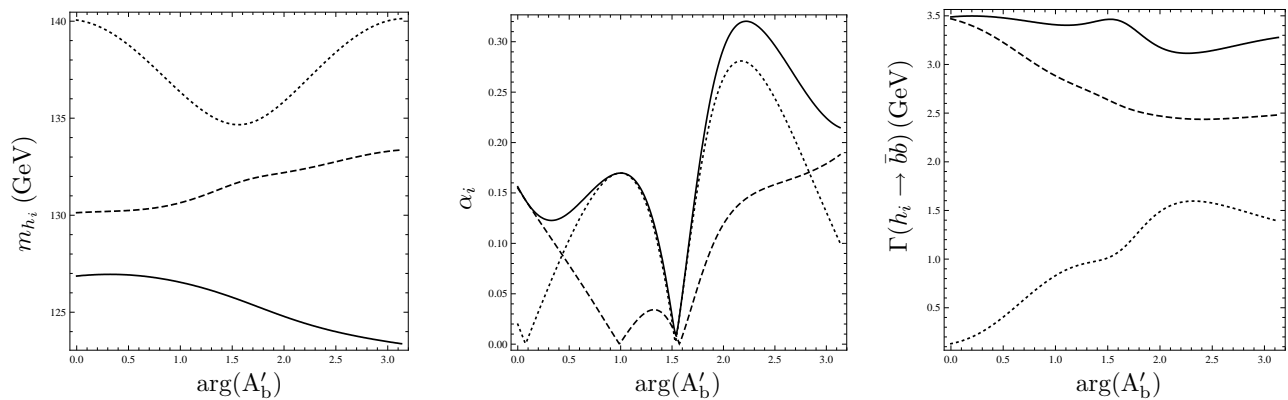}
\end{center}
\vspace{-0.4cm} \caption{The same with Fig. \ref{fig5} but now
$\tan\beta=50$, $M_A=130$ GeV,
 plots are presented against  varying $A^\prime_b$. Inputs: $A_t=A_b=A^\prime_t=|A^\prime_b|=2 m_{\tilde Q}$}
\label{fig6}
\end{figure}

In order to show the importance of $A^\prime_b$ contribution, we
present Fig. \ref{fig6}, in which  all the input parameters are the
same with Fig. \ref{fig5}, but now $\tan\beta$=50, $M_A=130$ GeV.
Here it is interesting to observe that the coupling of the lightest
neutral Higgs boson is very strong, sensitive to the argument of the
$A^\prime_b$, additionally $\Gamma(h_1\rightarrow\bar{b}b)$ can be
as large as $\geq$3 GeV (solid line of the right panel) which is
well above the SM prediction. A common property of the Figs.
\ref{fig5},\ref{fig6} is that when $\theta_{eff}\neq0$,  mass
difference of $h_2$ and $h_3$ bosons increases.

Notice that if were to assign $arg(A_f)=-arg(A^\prime_f)$ then there
would be no variation for the masses, the CP compositions and the
partial decay widths of Higgs bosons which can be important for CP
violating issues such as Electric Dipole Moments (EDMs) of
fundamental fermions. This can be seen from the definition of the CP
violating parameter $I_f$.

It can be inferred from the presented figures that the NH soft
breaking terms can yield sizable variations on the masses, CP
compositions and partial decay widths of neutral Higgs bosons. These
decay widths covers the range from $\sim$ SM values up to 3.5 GeV,
CP violating parameters can be obtained from zero to one, thanks to
the NH terms. Some of these results can also be simulated with a
complex $\mu$ parameter of the MSSM (when
$A^\prime_t=A^\prime_b=\mu$), but the option of replacing this
parameter with the non-holomorphic ones should be seen as  an
attractive alternative for the continuing and coming Higgs searches.
While not presented here, it is easy to guess that the couplings of
Higgs bosons to vector bosons  are also sensitive to the mentioned
NH terms.
\section{Conclusion}
In this paper, we studied the mass matrix of the neutral Higgs
bosons in the NHSSM with explicit CP violation at the one-loop
level. For doing this, we first obtained analytical expressions for
the mass matrix of neutral Higgs bosons (see the Appendix) and
performed a numerical study based on the new sources of CP violating
trilinear terms.

In order to maximize the impact of the NH terms, among many possible
parametrizations of the NH model, a special one is selected in which
the indirect relation between scalar fermions and inos over the
$\mu$ parameter is disappeared. In this version of the NH model,
$\mu$ term is absent in the superpotential and it exists in the soft
breaking part as a mass term for Higgsinos.  We observed that this
can heavily effect the mass and the mixings of the neutral Higgs
bosons of the MSSM.

During the numerical analysis we intentionally considered beyond the
MSSM scenarios (such as $A^\prime_t\,\neq\,A^\prime_b$) and observed
that not only holomorphic soft breaking terms but also
non-holomorphic terms can induce sizable amount of CP violation in
the Higgs sector. In order to show this we studied numerically
partial decay widths and CP-violating parameters of all the neutral
Higgs bosons.  We believe the selected ranges of our examples can be
important for the continuing and upcoming ($i.e.$ see
\cite{DeRoeck:2002hk}) Higgs searches with a generalized soft
breaking MSSM. Additionally, this issue should be probed deeper
because new sources of CP violating terms consisting with current
collider bounds can be useful, for instance, to relax the electron
and neutron EDM bounds on the CP violating terms of the MSSM
\cite{Abel:2001vy}. Analysis of various observables ranging from
$b\rightarrow\,s \gamma$ decay to EDM constrains  can shed further
light on the structure of non-holomorphic models.
\section{Acknowledgments}
A.S gratefully acknowledges the support from the Finnish Center for International Mobility (CIMO).

\appendix
\section{Definitions and Matrix Elements}
The definitions that appear in our
calculations are as follows:  For the proper treatment of the amount
of CP violation we collected imaginary and real parts of frequently
appearing terms as
$\mathcal{I}_b=\text{Im}(A_b\,A^\prime_b\,e^{i\theta})$,
$\mathcal{I}_t=\text{Im}(A_t\,A^\prime_t\,e^{i\theta})$, similarly
$\mathcal{R}_b=\text{Re}(A_b\,A^\prime_b\,e^{i\theta})$ and
$\mathcal{R}_t=\text{Re}(A_t\,A^\prime_t\,e^{i\theta})$. This
enables one to decompose scalar fermions into parts as \bea
&&\Sigma_B=2m_{\tilde{t}_L}^2+2m_{\tilde{b}_R}^2+v^2_u\Sigma_{G_b}+v^2_d\left(2|h_b|^2-\Sigma_{G_b}\right),\Delta_B=\kappa^2_1/\Delta^2_{G_b}+8|h_b|^2\left(|A_b|^2v^2_d+v_u\left(|A^\prime_b|^2v_u-2v_d\mathcal{R}_b\right)\right)\,\,\,\,\,\,\\
&&\Sigma_T=2m_{\tilde{t}_L}^2+2m_{\tilde{t}_R}^2-v^2_d{\Sigma_{G_t}}+v^2_u\left(2|h_t|^2+{\Sigma_{G_t}}\right),\Delta_T=\kappa^2_2/\Delta^2_{G_t}+8|h_t|^2\left(|A^\prime_t|^2v^2_d+v_u\left(|A_t|^2v_u-2v_d\mathcal{R}_t\right)\right)\,\,\,\,\,\,
\eea where \bea
&&\kappa_{1}={\Delta_{G_b}}\left(2m_{\tilde{b}_R}^2-2m_{\tilde{t}_L}^2+\left({v_d}^2-{v_u}^2\right){\Delta_{G_b}}\right),\,\,\,\kappa_{2}={\Delta_{G_t}}\left(-2m_{\tilde{t}_L}^2+2m_{\tilde{t}_R}^2+\left({v_d}^2-{v_u}^2\right){\Delta_{G_t}}\right)\\
&&\kappa_{3}={\Delta_{G_b}}\left(2m_{\tilde{b}_R}^2-2m_{\tilde{t}_L}^2+\left({v_d}^2-3{v_u}^2\right){\Delta_{G_b}}\right),\,\,\,\kappa_{4}={\Delta_{G_t}}\left(-2m_{\tilde{t}_L}^2+2m_{\tilde{t}_R}^2+\left({v_d}^2-3{v_u}^2\right){\Delta_{G_t}}\right)\\
&&\kappa_{5}={\Delta_{G_b}}\left(2m_{\tilde{b}_R}^2-2m_{\tilde{t}_L}^2+\left(3{v_d}^2-{v_u}^2\right){\Delta_{G_b}}\right),\,\,\,\kappa_{6}={\Delta_{G_t}}\left(-2m_{\tilde{t}_L}^2+2m_{\tilde{t}_R}^2+\left(3{v_d}^2-{v_u}^2\right){\Delta_{G_t}}\right)\\
&&\Delta_{G_t}=\left(-3g^2_2+5g^2_Y\right)/12,\,\,\Sigma_{G_t}=-\left(g^2_2+g^2_Y\right)/4,\,\,\Delta_{G_b}=\left(3g^2_2-g^2_Y\right)/12,\,\,\Sigma_{G_b}=\left(g^2_2+g^2_Y\right)/4\eea
Additionally, we defined the following quantities such that entries
of the Higgs matrix can be expressed in simpler forms. \bea
&&\chi_1=4\left(\left(4{A_b}^2|h_b|^2+\kappa_1\right){v_d}-4|h_b|^2{v_u}\mathcal{R}_b\right),\,\chi_2=4\left(\left(4{A^\prime_t}^2|h_t|^2+\kappa_2\right){v_d}-4|h_t|^2{v_u}\mathcal{R}_t\right),\,\chi_3=4\left(4{A_b}^2|h_b|^2+\kappa_5\right)\nonumber\\
&&\chi_4=4\left(4{A^\prime_t}^2|h_t|^2+\kappa_6\right),\,\chi_5=4\left(4{A^\prime_b}^2|h_b|^2{v_u}-\kappa_1{v_u}-4|h_b|^2{v_d}\mathcal{R}_b\right),\,\chi_6=4\left(4{A_t}^2|h_t|^2{v_u}-\kappa_2{v_u}-4|h_t|^2{v_d}\mathcal{R}_t\right)\nonumber\\
&&\chi_7=-8\left({v_d}{v_u}{\Delta^2_{G_b}}+2|h_b|^2\mathcal{R}_b\right),\,\chi_8=-8\left({v_d}{v_u}{\Delta^2_{G_t}}+2|h_t|^2\mathcal{R}_t\right),\,\chi_9=16{A^\prime_b}^2|h_b|^2-4\kappa_3,\,\chi_{10}=16{A_t}^2|h_t|^2-4\kappa_4.\nonumber
\eea
Using the definitions given above, the elements for the mass
matrix of the neutral Higgs bosons due to radiative contribution of
quarks and squarks are obtained as follows: \bea
&&\mathcal{M}^2_{11}=\frac{{\hat{g}}^{2}{v_d}^2}{4}+{m^2_3}\frac{{v_u}\cos[\theta+\Phi]}{{v_d}}+\frac{3\left(8{v_d}^2{\chi_2}\Delta_T{\Sigma_{G_t}}+2{\chi_2}\Delta_T\Sigma_T+{v_d}\left({\chi_2}^2-2{\chi_4}\Delta_T\right)\Sigma_T\right)}{1024\pi^2{v_d}\Delta_T^{3/2}}\ln\frac{m^2_{\tilde{t}_1}}{m^2_{\tilde{t}_2}}\\
&&-\frac{3|h_b|^4{v_d}^2}{8\pi^2}\ln\frac{m^2_b}{\Lambda^2}+\frac{3\left(2({\chi_1}+{\chi_2})\Delta_B\Delta_T+{v_d}\left({\chi_2}^2\Delta_B+\left({\chi_1}^2-2({\chi_3}+{\chi_4})\Delta_B\right)\Delta_T\right)\right)}{512\pi^2{v_d}\Delta_B\Delta_T}\nonumber\\
&&+\frac{3\left(2{\chi_1}\Delta_B+{v_d}\left({\chi_1}^2-2{\chi_3}\Delta_B\right)\right)\Sigma_B+24{v_d}^2{\chi_1}\Delta_B\left(-2|h_b|^2+{\Sigma_{G_b}}\right)}{1024\pi^2{v_d}\Delta_B^{3/2}}\ln\frac{m^2_{\tilde{b}_1}}{m^2_{\tilde{b}_2}}\nonumber\\
&&+\frac{3\left({v_d}^3\left({\Delta^2_{G_t}}+{\Sigma^2_{G_t}}\right)+2|h_t|^2{v_u}\mathcal{R}_t\right)}{64\pi^2{v_d}}\ln\frac{m^2_{\tilde{t}_1}m^2_{\tilde{t}_2}}{\Lambda^4}+\frac{3{v_d}^3\left({\Delta^2_{G_b}}+\left(-2|h_b|^2+{\Sigma_{G_b}}\right)^2\right)+6|h_b|^2{v_u}\mathcal{R}_b}{64\pi^2{v_d}}\ln\frac{m^2_{\tilde{b}_1}m^2_{\tilde{b}_2}}{\Lambda^4},\nonumber\eea

\bea
&&\mathcal{M}^2_{12}=-\frac{1}{4}{\hat{g}}^{2}{v_d}{v_u}+{m^2_3}(-\cos[\theta+\Phi])+\frac{3({\chi_2}{\chi_6}\Delta_B+({\chi_1}{\chi_5}-2({\chi_7}+{\chi_8})\Delta_B)\Delta_T)}{512\pi^2\Delta_B\Delta_T}\\
&&+\frac{3\left(-8|h_t|^2{v_u}{\chi_2}\Delta_T-4{v_u}{\chi_2}\Delta_T{\Sigma_{G_t}}+4{v_d}{\chi_6}\Delta_T{\Sigma_{G_t}}+{\chi_2}{\chi_6}\Sigma_T-2{\chi_8}\Delta_T\Sigma_T\right)}{1024\pi^2\Delta_T^{3/2}}\ln\frac{m^2_{\tilde{t}_1}}{m^2_{\tilde{t}_2}}\nonumber\\
&&+\frac{3\left(-8|h_b|^2{v_d}{\chi_5}\Delta_B+{\chi_1}{\chi_5}\Sigma_B-2{\chi_7}\Delta_B\Sigma_B-4{v_u}{\chi_1}\Delta_B{\Sigma_{G_b}}+4{v_d}{\chi_5}\Delta_B{\Sigma_{G_b}}\right)}{1024\pi^2\Delta_B^{3/2}}\ln\frac{m^2_{\tilde{b}_1}}{m^2_{\tilde{b}_2}}\nonumber\\
&&+\frac{3\left({\chi_8}-8{v_d}{v_u}{\Sigma_{G_t}}\left(2|h_t|^2+{\Sigma_{G_t}}\right)\right)}{512\pi^2}\ln\frac{m^2_{\tilde{t}_1}m^2_{\tilde{t}_2}}{\Lambda^4}+\frac{3\left({\chi_7}-8{v_d}{v_u}{\Sigma_{G_b}}\left(-2|h_b|^2+{\Sigma_{G_b}}\right)\right)}{512\pi^2}\ln\frac{m^2_{\tilde{b}_1}m^2_{\tilde{b}_2}}{\Lambda^4},\nonumber
\eea

\bea
&&\mathcal{M}^2_{13}={m^2_3}\frac{{v_d}\text{Sin}[\theta+\Phi]}{v}+\frac{3\left(|h_b|^2\left(v^2{\chi_1}-2{v_d}\Delta_B\right)\Delta_T\mathcal{I}_b+|h_t|^2\Delta_B\left(v^2{\chi_2}-2{v_d}\Delta_T\right)\mathcal{I}_t\right)}{32\pi^2v\Delta_B\Delta_T}\\
&&+\frac{3|h_t|^2{v_d}\mathcal{I}_t}{32\pi^2v}\ln\frac{m^2_{\tilde{t}_1}m^2_{\tilde{t}_2}}{\Lambda^4}+\frac{3|h_t|^2\left(-2{v_d}\Delta_T\Sigma_T+v^2(4{v_d}\Delta_T{\Sigma_{G_t}}+{\chi_2}\Sigma_T)\right)\mathcal{I}_t}{64\pi^2v\Delta_T^{3/2}}\ln\frac{m^2_{\tilde{t}_1}}{m^2_{\tilde{t}_2}}\nonumber\\
&&+\frac{3|h_b|^2{v_d}\mathcal{I}_b}{32\pi^2v}\ln\frac{m^2_{\tilde{b}_1}m^2_{\tilde{b}_2}}{\Lambda^4}-\frac{3|h_b|^2\left(-v^2{\chi_1}\Sigma_B+2{v_d}\Delta_B\Sigma_B-4v^2{v_d}\Delta_B\left(-2|h_b|^2+{\Sigma_{G_b}}\right)\right)\mathcal{I}_b}{64\pi^2v\Delta_B^{3/2}}\ln\frac{m^2_{\tilde{b}_1}}{m^2_{\tilde{b}_2}},\nonumber
\eea

\bea
& &\mathcal{M}^2_{23}={m^2_3}\frac{{v_u}\text{Sin}[\theta+\Phi]}{v}+\frac{3\left(|h_b|^2\left(v^2{\chi_5}-2{v_u}\Delta_B\right)\Delta_T\mathcal{I}_b+|h_t|^2\Delta_B\left(v^2{\chi_6}-2{v_u}\Delta_T\right)\mathcal{I}_t\right)}{32\pi^2v\Delta_B\Delta_T}\\
&&-\frac{3|h_t|^2\left(4v^2{v_u}\Delta_T\left(2|h_t|^2+{\Sigma_{G_t}}\right)-v^2{\chi_6}\Sigma_T+2{v_u}\Delta_T\Sigma_T\right)\mathcal{I}_t}{64\pi^2v\Delta_T^{3/2}}\ln\frac{m^2_{\tilde{t}_1}}{m^2_{\tilde{t}_2}}+\frac{3|h_t|^2{v_u}\mathcal{I}_t}{32\pi^2v}\ln\frac{m^2_{\tilde{t}_1}m^2_{\tilde{t}_2}}{\Lambda^4}\nonumber\\
&&+\frac{3|h_b|^2\left(-2{v_u}\Delta_B\Sigma_B+v^2({\chi_5}\Sigma_B-4{v_u}\Delta_B{\Sigma_{G_b}})\right)\mathcal{I}_b}{64\pi^2v\Delta_B^{3/2}}\ln\frac{m^2_{\tilde{b}_1}}{m^2_{\tilde{b}_2}}+\frac{3|h_b|^2{v_u}\mathcal{I}_b}{32\pi^2v}\ln\frac{m^2_{\tilde{b}_1}m^2_{\tilde{b}_2}}{\Lambda^4},\nonumber\eea

\bea
&&\mathcal{M}^2_{22}=\frac{{\hat{g}}^{2}{v_u}^2}{4}+{m^2_3}\frac{{v_d}\cos[\theta+\Phi]}{{v_u}}+\frac{3\left(2({\chi_5}+{\chi_6})\Delta_B\Delta_T+{v_u}\left({\chi_6}^2\Delta_B+\left({\chi_5}^2-2({\chi_{10}}+{\chi_9})\Delta_B\right)\Delta_T\right)\right)}{512\pi^2{v_u}\Delta_B\Delta_T}\\
&&-\frac{3|h_t|^4{v_u}^2}{8\pi^2}\ln\frac{m^2_t}{\Lambda^2}+\frac{3\left(2{\chi_5}\Delta_B\Sigma_B+{v_u}\left({\chi_5}^2-2{\chi_9}\Delta_B\right)\Sigma_B-8{v_u}^2{\chi_5}\Delta_B{\Sigma_{G_b}}\right)}{1024\pi^2{v_u}\Delta_B^{3/2}}\ln\frac{m^2_{\tilde{b}_1}}{m^2_{\tilde{b}_2}}\nonumber\\
&&+\frac{-24{v_u}^2{\chi_6}\Delta_T\left(2|h_t|^2+{\Sigma_{G_t}}\right)+3\left(2{\chi_6}\Delta_T+{v_u}\left({\chi_6}^2-2{\chi_{10}}\Delta_T\right)\right)\Sigma_T}{1024\pi^2{v_u}\Delta_T^{3/2}}\ln\frac{m^2_{\tilde{t}_1}}{m^2_{\tilde{t}_2}}\nonumber\\
&&+\frac{3{v_u}^3\left({\Delta^2_{G_t}}+\left(2|h_t|^2+{\Sigma_{G_t}}\right)^2\right)+6|h_t|^2{v_d}\mathcal{R}_t}{64\pi^2{v_u}}\ln\frac{m^2_{\tilde{t}_1}m^2_{\tilde{t}_2}}{\Lambda^4}+\frac{3\left({v_u}^3\left({\Delta^2_{G_b}}+{\Sigma^2_{G_b}}\right)+2|h_b|^2{v_d}\mathcal{R}_b\right)}{64\pi^2{v_u}}\ln\frac{m^2_{\tilde{b}_1}m^2_{\tilde{b}_2}}{\Lambda^4}.\nonumber
\eea

\end{document}